\documentclass[twocolumn]{aastex62}
 
\usepackage{graphicx}
\usepackage{subfigure}
\usepackage{multirow}
\usepackage{comment}
\usepackage{natbib}
\usepackage{hyperref}
\usepackage{mathtools}
\usepackage{mathrsfs}
\usepackage{fontenc}
\usepackage{color}
\usepackage{url}
\usepackage{hyperref}
\usepackage{pifont}
\usepackage{textcomp}
\usepackage{color}
\usepackage{multirow}
\usepackage{booktabs}
\usepackage{array}

\def\apjl{ApJL}
\def\apjs{ApJS}
\def\ao{Appl.~Opt.}
\def\aap{A\&A}
 
\DeclareMathAlphabet{\mathsc}{OT1}{cmr}{m}{sc}
\def\testbx{bx}%
\DeclareRobustCommand{\ion}[2]{%
\relax\ifmmode
\ifx\testbx\f@series
{\mathbf{#1\,\mathsc{#2}}}\else
{\mathrm{#1\,\mathsc{#2}}}\fi
\else\textup{#1\,{\mdseries\textsc{#2}}}%
\fi}

\newcommand{\kms}{\mbox{$\rm{\,km\,s^{-1}}$}}

\def\lax{{$\mathrel{\hbox{\rlap{\hbox{\lower4pt\hbox{$\sim$}}}\hbox{$<$}}}$}}
\def\gax{{$\mathrel{\hbox{\rlap{\hbox{\lower4pt\hbox{$\sim$}}}\hbox{$>$}}}$}}

\def\cmss{cm s$^{-2}$}

\def\mgii{Mg~{\sc ii}~}

\def\civ{C~{\sc iv}~}
\def\aliii{Al~{\sc iii}~}
\def\siv{Si~{\sc iv}~}
\def\nv{N~{\sc v}}

\def\zem{$z_{\rm em}$~} 
 
\def\lya{Ly$\alpha$~}

\def\kms{km s$^{-1}$} 
\def\apjl{{ApJ}}%
\def\apjs{{ApJS}}%
\def\aap{{A\&A}}%
\begin{document}

\title{Deceleration of \civ and \siv  broad absorption lines in X-ray bright quasar SDSS-J092345$+$512710}

\shorttitle{BAL deceleration}

\shortauthors{JOSHI ET AL}

\author{Ravi Joshi}
\affil{Kavli Institute for Astronomy and Astrophysics, Peking University, Beijing 100871, China}

\author{Raghunathan Srianand}
\affil{Inter-University Centre for Astronomy and Astrophysics, Post Bag 4, Ganeshkhind, Pune 411007, India }

\author{Hum Chand}
\affil{Aryabhatta Research Institute of Observational Sciences (ARIES), Manora Peak, Nainital $-$ 263 002, India}

\author{Xue-Bing Wu}
\affil{Kavli Institute for Astronomy and Astrophysics, Peking University,
Beijing 100871, China}
\affil{Department of Astronomy, School of Physics, Peking University,
Beijing 100871, China}

\author{Pasquier Noterdaeme}
\affil{Institut d'Astrophysique de Paris, UMR 7095, CNRS-SU, 98bis bd Arago, 75014 Paris, France }

\author{Patrick  Petitjean}
\affil{Institut d'Astrophysique de Paris, UMR 7095, CNRS-SU, 98bis bd Arago, 75014 Paris, France }

\author{Luis C.  Ho}
\affil{Kavli Institute for Astronomy and Astrophysics, Peking University,
Beijing 100871, China}
\affil{Department of Astronomy, School of Physics, Peking University,
Beijing 100871, China}

\correspondingauthor{Ravi Joshi}
\email{rvjoshirv@gmail.com}

\begin{abstract}
We report a synchronized kinematic shift of \civ and \siv broad
absorption lines (BAL) in a high-ionization, radio-loud, and X-ray
bright quasar SDSS-J092345$+$512710 (at \zem\ $\sim$ 2.1627). This
quasar shows two broad absorption components (blue component at $v
\sim 14,000$~\kms, and red component at $v \sim 4,000$~\kms\ with
respect to the quasars systemic redshift). The absorption profiles of
\civ and \siv BAL of the blue component show decrease in outflow
velocity with an average deceleration rate of $
-1.62_{-0.05}^{+0.04}$~\cmss\ and $-1.14^{+0.21}_{-0.22}$~\cmss\ over
a rest-frame time-span of 4.15 years. We do not see any 
  acceleration-like signature in the red component. This is
consistent with dramatic variabilities usually seen at high
velocities. During our monitoring period the quasar has shown no
strong continuum variability. We suggest the observed variability
could be related to the time dependent changes in disk wind parameters
like launching radius, initial flow velocity or mass outflow rate.

\end{abstract}

\section{Introduction}
\label{lab:xbal_intro}

Outflows are ubiquitous and appear to be the main  source  of active
galactic nucleus (AGN) feedback which regulates black hole growth and
host galaxy evolution as well as enrich the intergalactic and
circumgalactic medium around galaxies
\citep[see][]{Ostriker2010ApJ...722..642O,Kormendy2013ARA&A..51..511K}.
These feedback processes most likely drive the well-known observed
correlation between the supermassive black hole (SMBH) mass and
physical properties of the host galaxy, along with the steep decline
in the number density of galaxies at high masses
\citep{Ferrarese2000ApJ...539L...9F,Hopkins2005ApJ...630..705H,
  Ostriker2010ApJ...722..642O}. The signatures of strong outflows are
directly observed in roughly 20 percent of quasar population via broad
ultraviolet-resonance absorption lines (BAL)
\citep[][]{Weymann1991ApJ...373...23W,
  Trump2006ApJS..165....1T,Gibson2009ApJ...692..758G}, spanning a
large range of outflow velocities from $1000$~\kms\ up to several
$10,000$
\kms\ \citep[e.g.,][]{Weymann1991ApJ...373...23W,Hamann1997ApJ...478...87H,Hidalgo2011MNRAS.411..247R,Rogerson2016MNRAS.457..405R}.
Nonetheless, many aspects of quasar outflows remain poorly understood,
including the gas geometry, acceleration mechanism(s) and their
influence on the host galaxy and its environments. \par

BAL variability study is a promising technique for constraining the
structure and location of the associated wind. In recent systematic
studies of BAL variability, BAL troughs are commonly observed to show variability
in absorption strength
\citep{Barlow1994PASP..106..548B,Lundgren2007ApJ...656...73L,Gibson2008ApJ...685..773G,Capellupo2011MNRAS.413..908C,Capellupo2012MNRAS.422.3249C,Capellupo2013MNRAS.429.1872C,Ak2013ApJ...777..168F,Vivek2014MNRAS.440..799V,Grier2015ApJ...806..111G,McGraw2018MNRAS.475..585M}
and/or profile, also known as ``transient BALs'' showing emergence or
disappearance of BAL features
\citep[e.g.,][]{Hamann2008MNRAS.391L..39H,Hall2011MNRAS.411.2653H,
  Hidalgo2011MNRAS.411..247R,Ak2012ApJ...757..114F,Vivek2016MNRAS.455..136V,
  McGraw2017MNRAS.469.3163M}, over a broad range of rest-frame
timescales, ranging from months to years. However, the signature of
acceleration (e.g., kinematic shift of absorption profile) are more
scarce, reported only a few times
\citep[e.g.,][]{Vilkoviskij2001MNRAS.321....4V,
  Rupke2002ApJ...570..588R, Gabel2003ApJ...595..120G,
  Hall2007ApJ...665..174H,Joshi2014MNRAS.442..862J,
  Grier2016ApJ...824..130G}. Mechanisms proposed to understand the
observed BAL variability invoke  changes in ionization
state and/or the movement of individual clouds or substructures in the
outflow, across our line of sight
\citep{Lundgren2007ApJ...656...73L,Hall2007ApJ...665..174H,Hamann2008MNRAS.391L..39H}.
In most cases the BAL variations are not found to be correlated with
the optical continuum variations. Also not all velocity components
seen in absorption show correlated variations. This hints towards
mechanisms other than photoionization induced variations. However, a
unified understanding of BAL variations is an ongoing endeavor.

 \par

The BAL features are widely believed to be formed in ``disk winds'',
launched from the surface of the accretion disk at 10-100 light days
from the central SMBH (of $\sim 10^9 \rm \ M\odot$) mainly driven by
radiative forces
\citep{Arav1994ApJ...427..700A,Murray1995ApJ...451..498M,
  Proga2000ApJ...543..686P,Higginbottom2014ApJ...789...19H}. In
radiation-driven scenarios, the wind is efficiently accelerated to
high velocities by invoking the shielding gas close to the base of
outflow which prevents the UV-absorbing gas from becoming overionized
by nuclear X-ray and extreme-ultraviolet (UV) photons
\citep{Murray1995ApJ...451..498M,Proga2000ApJ...543..686P}. The above
paradigm is challenged by the observed flows having high velocities
and lower degree of ionization in X-ray bright mini-BAL (typical full
width half maximum of $500 - 2000$~\kms) quasars
\citep{Hamann2008MNRAS.391L..39H,Hamann2013MNRAS.435..133H}. These
observations favor the substructured flow, involving tiny dense clouds
with a low volume filling factor, driven out by radiative forces while
being confined by magnetic pressure
~\citep{Rees1987QJRAS..28..197R,Baskin2014MNRAS.445.3025B,Matthews2016MNRAS.458..293M}.
It suggests that the strong radiative shielding gas may not be
universal component of quasar outflows for accelerating the gas to
high speeds. This idea is also supported by the recent high-energy
X-ray observations showing that a large fraction, $\sim 6-23\%$, of
BAL quasars among the general BAL quasar population are perhaps
intrinsically X-ray weak in nature
\citep{Luo2013ApJ...772..153L,Luo2014ApJ...794...70L,Teng2014ApJ...785...19T,Liu2018ApJ...859..113L}.

The emerging picture of BAL outflows suggests that the mini-BALs and
BALs arise from the same quasar wind, where BALs form in the main part
of the outflow near the accretion disc plane while mini-BALs form
along sightlines at higher latitudes
\citep{Ganguly2001ApJ...549..133G,Hamann2008MNRAS.391L..39H,Hamann2013MNRAS.435..133H}.
This also explains the observed X-ray bright nature of mini-BALs.
Interestingly, a new population of X-ray bright BAL quasars is
recently discovered in X-ray surveys
\citep[e.g.,][]{Giustini2008A&A...491..425G,Gibson2009ApJ...692..758G,
  Streblyanska2010AIPC.1248..513S,Liu2018ApJ...859..113L} which
further possess major challenges to the models of BAL outflows. Note
that, if the X-ray bright BAL quasars are preferentially originated
in a structured flow viewed along the sight lines of higher latitudes
than one would naively expect to see the combination of line shift and
line strength variability in this subclass. Indeed, in our recent
efforts to probe the variability nature of X-ray bright BAL quasars we
have find two such rare cases of kinematic shift and strength
variability of the \civ BAL trough \citep[e.g.,][]{Joshi2014MNRAS.442..862J}. Given the X-ray weak nature of
general population of BAL quasars, a systematic study of this rare
population of X-ray bright BAL quasars will provide important
observational constrains on the BAL geometry and the physical
mechanisms for launching and accelerating the quasar outflows.

Here, we report the detection of a  deceleration-like signature
  in \civ and \siv BAL outflows towards X-ray bright quasar
J092345$+$512710. This source is part of our on going monitoring
program of BAL spectral variability in rare X-ray bright BAL quasars
\citep[see,][]{Joshi2014MNRAS.442..862J}. This paper is organized as
follows. Section~\ref{lab:xbal_obsand dataredu} describes the
observations and data reduction. In Section~\ref{lab:xbal_RES}, we
present the results of our analysis followed with the discussion and
conclusion in Section~\ref{lab:xbal_DnC}.

\section{Observation and Data Reduction}
\label{lab:xbal_obsand dataredu}

 The BAL quasar J092345$+$512710 (\zem = 2.1627) was first detected in Sloan Digital
 Sky Survey \citep{Trump2006ApJS..165....1T} which we have followed
 with 2-metre telescope at IUCAA Girawali Observatory (IGO), using
 IUCAA Faint Object Spectrograph and Camera (IFOSC). We performed a
 long-slit spectroscopic observation of this target on 2011 April 1.
 In order to cover the \civ and \siv BAL troughs we have used
 Grism7 of IFOSC, in combination with a 1.5 arcsec slit. This yielded a
 wavelength coverage of 3800$-$6840~\AA\ at a spectral resolution $R
 \sim$~1140 (i.e., $\sim$ 310~\kms). We acquired 2 exposures of 45
 minutes each. The raw CCD frames were cleaned using standard {\sc
   iraf}{\footnote{\textsc{iraf} is distributed by the
     \textsc{National Optical Astronomy Observatories}, which are
     operated by the Association of Universities for Research in
     Astronomy, Inc., under cooperative agreement with the National
     Science Foundation.}} procedures. We carried out the bias and
 flat-field corrections to all the frames. The Halogen flats were used
 for flat fielding the frames. We then extracted the
 one-dimensional spectrum from individual frames using the
 \textsc{iraf} task ``apall''. Wavelength calibration was done using
 the standard Helium-Neon lamp spectra and flux calibration was done
 using standard star observed on the same night. We applied
 air-to-vacuum conversion and coadded the spectra, using
 ${1}/{\sigma_i^2}$ weighting in each pixel, after scaling the overall
 individual spectrum to a common flux level. Subsequently the quasar
 was observed twice again as part of SDSS-BOSS survey. This four epochs
 data forms the main resource for the analysis presented here.
 Details are given in Table~\ref{lab:tab_sourceobs_info}.

 We also used photometric light curves from publicly available Catalina Real-Time Transient
Survey{\footnote{http://nunuku.cacr.caltech.edu/cgi-bin/getcssconedb\_release\_img.cgi}}
(CRTS) to judge  the amount of flux variations seen in this quasar.

\begin{table*}
 \centering
 \begin{minipage}{155mm}
{\small
\caption{Information related to observations and  other basic  parameters of J092345$+$512710 spectra and its  BAL trough.}
\label{lab:tab_sourceobs_info}
\begin{tabular}{@{}clc ccrrccc@{}} 
\hline 
 \multicolumn{1}{l}{Instrument}  
& \multicolumn{1}{c}{Date (Epoch,yy.mm.dd)} 
& \multicolumn{1}{c}{Exposure}
 & \multicolumn{1}{c}{Resolution}
& \multicolumn{1}{c}{S/N$^{\textcolor{blue}{a}}$ }
& \multicolumn{1}{c}{$W_{1549}$$^{\textcolor{blue}{b}}$}
& \multicolumn{1}{c}{$W_{1400}$}
& \multicolumn{1}{c}{$W_{1549}$/$W_{1400}$}
\\
&\multicolumn{1}{c}{(MJD)}
&\multicolumn{1}{c}{Time (mins)}
&\multicolumn{1}{c}{(\kms)}
&
& \multicolumn{1}{c}{(\AA)}
& \multicolumn{1}{c}{(\AA)}
&
\\
\hline 
SDSS                                        & 52252 (\#1, 2001.09.12) &80$\times$1    &150 & 11 &$14.0\pm0.5$  &$ 9.4\pm0.5$ & 1.5$\pm$0.1\\
{IGO/IFOSC 7$^{\textcolor{blue}{c}}$}         & 55652 (\#2, 2011.07.01) &45$\times$2    &310  &  4 &$17.7\pm1.3$ &$11.8\pm1.1$ & 1.5$\pm$0.2\\
SDSS-BOSS                                   & 56607 (\#3, 2013.11.11) &75$\times$1    &150 & 14 &$20.8\pm0.3$  &$ 8.8\pm0.3$ & 2.5$\pm$0.1 \\
SDSS-BOSS                                   & 57046 (\#4, 2015.24.01) &90$\times$1    &150 & 11 &$19.1\pm0.4$  &$ 6.1\pm0.4$ & 3.1$\pm$0.2\\
\hline   
\end{tabular}                              
}
\\                          
$^a$ Signal-to-noise ratio per-pixel over the wavelength range 5100$-$5500 \AA.\\
$^b$ $W_{1549}$ is measured over a velocity range of -22,576 to -8,735 \kms. \\
$^c$ Wavelength Coverage of 3800$-$6840~\AA. 

\end{minipage}    

\end{table*}

\subsection{Continuum Fitting}
Following \citet{Gibson2009ApJ...692..758G}, we model the quasar
continuum emission using a Small Magellanic Cloud-like reddened
power-law function from \citet{Pei1992ApJ...395..130P}. We use the
emission redshift of \zem$= 2.16274$, from
~\citet{Hewett2010MNRAS.405.2302H} to get the rest frame quasar
spectrum and fit only regions largely free from emission and
absorption features which are in ranges $1280-1300$~\AA,
$1700-1800$~\AA, $1950-2200$~\AA, $2650-2710$~\AA, and $2950-3700$~\AA.
To the uncertainties over the continuum fit, we performed Monte
Carlo simulations by randomizing the flux in each pixel with a random
Gaussian deviate associated with uncertainty over the pixel. We fit
the continuum to the new spectrum over 100 times and adopt their
standard deviation as the uncertainties of the continuum fit.

\begin{figure*}[!ht]
\centering
	\includegraphics[width=16.5cm]{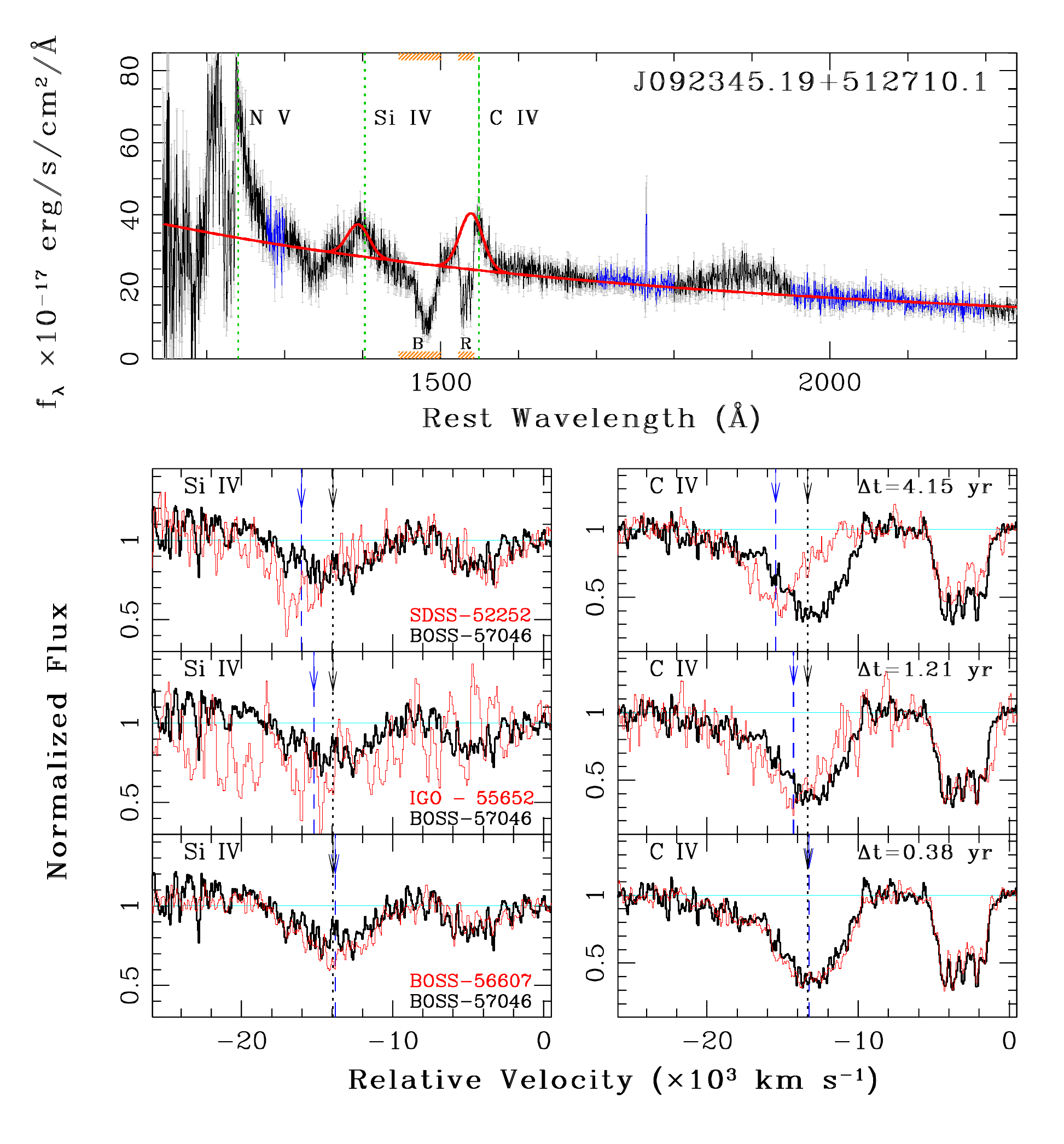}
   \caption{{Upper panel:} The final continuum fit (smooth curve)
     comprising of a power law and the multiple-Gaussian components
     for BOSS MJD-57046 (Epoch 4) spectrum. The hatched region show
     the ``blue'' and ``red'' \civ absorption profiles and marked as
     `B' and `R', respectively. The blue data intervals represent the
     relatively line-free windows chosen for the power-law fit. {Lower
       Panel:} The comparison of continuum-normalized \civ (right sub
     panels) and \siv (left sub panels) absorption line profiles
     (smoothed over 3 pixels) in SDSS MJD-52252 (Epoch 1), IGO
     MJD-55652 (Epoch 2), and BOSS MJD-56607 (Epoch 3) spectrum (red,
     thin solid line) with reference BOSS MJD-57046 (Epoch 4) spectrum
     (black, thick solid line). The velocity scale, with $ v =0 $
     \kms\ corresponding to quasar emission redshift of \zem\ $=
     2.1627$. The velocity centroid of absorption trough for reference
     (Epoch 4) and comparison spectrum are shown with arrow and also
     with dotted and dashed lines, respectively.}
\label{fig:xbal_vary_0911}
\end{figure*}

In addition, to remove the emission line features we model them using
Gaussian profiles over continuum subtracted spectra, without
associating  any physical meaning. Note that, modelling the \civ
emission is nontrivial as the line profile is usually asymmetric and
blanket with multiple strong absorption features. Therefore, to model
the \civ emission profile we exclude the wavelength ranges  having
absorption signature and fit the remaining \civ feature using a double Gaussian
profile. We model the \siv emission line with a single Gaussian. The
final continuum fit (solid line), comprising of a power law for the line free continuum and the
multi-Gaussian components for the broad emission lines is shown in top panel of
Fig.~\ref{fig:xbal_vary_0911}. The flux uncertainties from both the
continuum fit and flux measurement errors  were propagated to determine the final
uncertainty on the normalized spectrum. Finally, we generate the
continuum-normalized spectra to examine the BAL variability.

\section{Analysis} 
\label{lab:xbal_RES}

J092345$+$512710 is a HiBAL quasar having smooth \civ and \siv BAL
profiles at a similar outflow velocity of $v \sim -9000 $ to  $-20,000$ \kms
(see, top panel of Fig.~\ref{fig:xbal_vary_0911}). We refer to this as
``blue'' component (marked as `B' in Fig~\ref{fig:xbal_vary_0911}). We
also detect \nv\ absorption corresponding to that of \civ BAL trough
in the BOSS spectra of Epoch 3 and 4 at very blue part of the spectra
having poor S/N. There is a hint of weak \lya absorption at a similar
velocity as the \civ BAL. In addition, we also detect associated \civ
and \nv\ absorption component with $v \sim -1000$ to $-6000$~\kms\ composed
of several narrow components. We refer to this as ``red'' component
(marked as `R' in Fig~\ref{fig:xbal_vary_0911}). The \siv is found to
be weak in this component. We also searched for the additional BAL
features in the spectrum but there are no \aliii and \mgii BALs at the
corresponding position of \civ BAL trough.

In Fig.~\ref{fig:xbal_vary_0911}, lower sub-panels, we compare the
continuum normalized \civ and \siv absorption line profiles (smoothed
over 3 pixels) in IGO spectrum obtained on MJD-55652 (Epoch 2) and SDSS spectrum
from epoch MJD-52252 (Epoch 1), MJD-56607 (Epoch 3) with a reference
BOSS spectrum obtained on MJD-57046 (i.e., year 2015), in velocity
scale with $ v =0 $ \kms\ corresponding to systemic redshift of the
quasar, i.e., \zem\ $= 2.1627$. A decrease in radial velocity for both
\civ and \siv BAL troughs in the ``blue'' component is apparent in
high signal-to-noise (S/N) ratio spectra between Epochs 1 and 4, over
a rest frame time scale of 4.15 years. Interestingly, a clear
signature of velocity shift for the \civ component is also seen in our
low S/N IGO spectrum (see, left sub panels of
Fig.~\ref{fig:xbal_vary_0911}), over a rest-frame time span of 2.95
years and 1.21 years between Epoch 1 and 2 and Epoch 2 and 4,
respectively. However, no such kinematic shift is seen between Epoch 3
and 4 over relatively shorter rest frame time span of $\sim$0.38
years. Additionally, for the ``red'' component that shows
narrow sub-structures  we do not find any kinematic shift which ensures that
the observed kinematic shift in the ``blue'' BAL trough is real. We
note that our IGO (Epoch 2) spectrum is too noisy at shorter
wavelengths where \siv absorption is present, hence, we do not use it
further to study the \siv BAL variability. Unfortunately, as the
\nv\ region is not covered in the early epoch SDSS and IGO spectrum
that prevents us from studying the kinematic shift of \nv.

\begin{figure}[!ht]
 	\centering
 	\includegraphics[width=\linewidth]{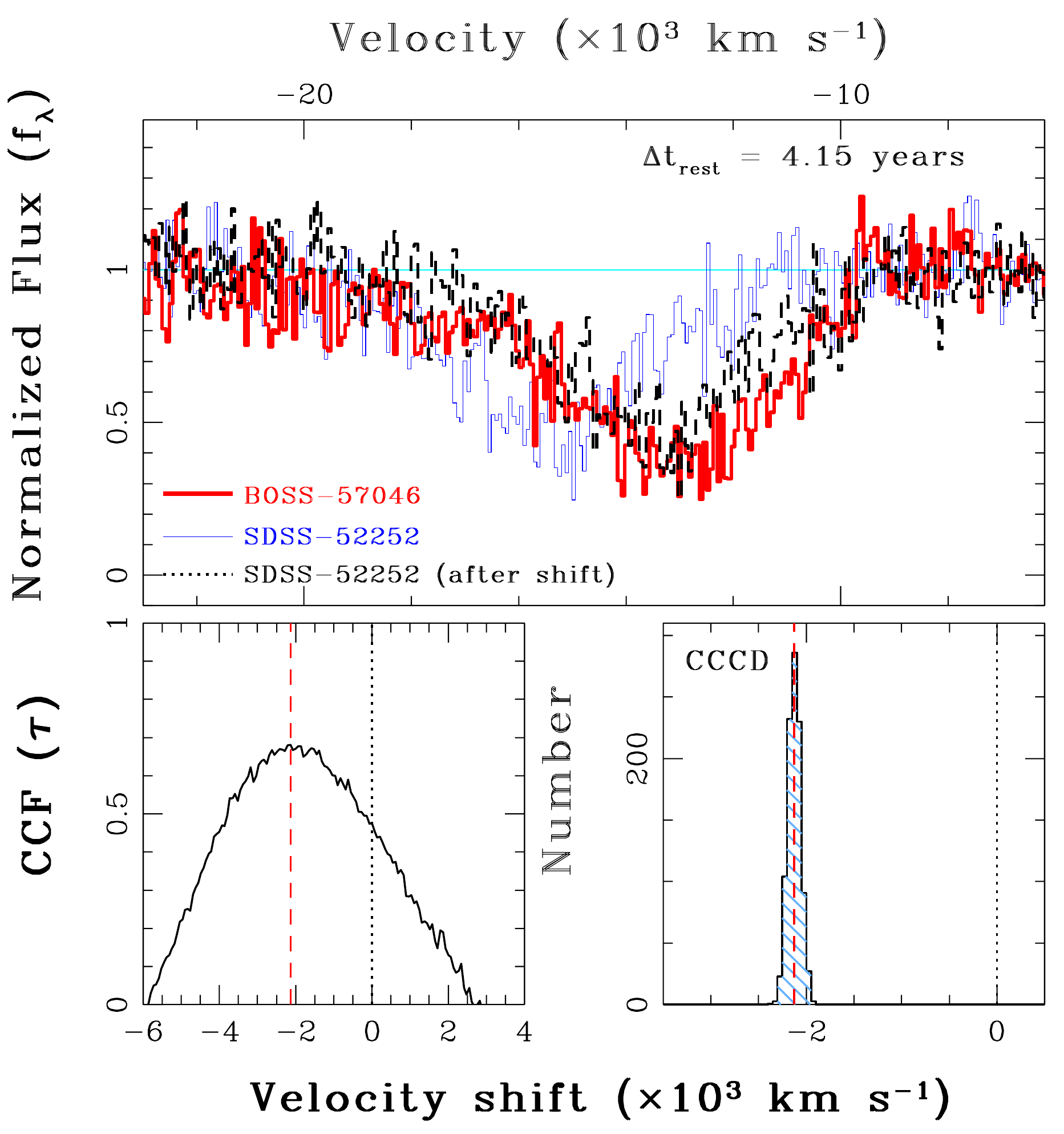}\\
 	\caption{ Top Panel: The comparison of \civ BAL profile between SDSS
   (Epoch 1) and BOSS (Epoch 4) spectra. The SDSS spectrum before and
   after shifting by the measured velocity shift from CCF analysis is
   shown in thin solid line (blue) and dotted line, respectively. Bottom
   Panel: The cross-correlation function (CCF) for the Epoch 1 and 4 spectra and
   cross correlation centroid (CCCD) distribution  from Monte-Carlo simulations are
   given in bottom-left and bottom-right panels, respectively. The
   centroid shift of BAL profile is given in dashed line with zero
   velocity point in dotted line.}
 	\label{fig:pixshift}
 \end{figure}

To quantify the kinematic shift of BAL trough, specifically blue
component, between two epochs spectra we perform a cross-correlation
function (CCF) analysis \citep[see also,][]{Grier2016ApJ...824..130G}.
For this, we consider the spectral region including the BAL trough in
question plus 2000 \kms\ on each side and measure the
cross-correlation coefficient ($\tau$) by shifting the early epoch
spectrum with 1 pixel step (i.e., 69 \kms) over a velocity range of
$-6000$ to +6000 \kms. We measure the velocity shift as the most
significant peak in the  correlation coefficient and the centroid of CCF
using only points around the peak (i.e, with $\tau > $ 0.8 $r_{\rm
  peak}$). In order to account for the measurement uncertainties we
perform the Monte Carlo simulations, randomizing the fluxes of both
spectra by a random Gaussian deviate associated with uncertainty over
each pixel. We generate 1000 such realizations and measure the CCF and
the corresponding peak and centroid for each realization of spectra.
We take the median of the cross correlation centroid distribution
(CCCD) as final velocity shift and 1$\sigma$ uncertainty as the
central interval encompassing 68\% of CCCD. The measured CCF and cross
correlation centroid distribution for two spectra from Epoch 1 and 4
are shown in lower-left and lower-right panels of
Fig.~\ref{fig:pixshift}. The best velocity shift and the corresponding
deceleration values between different epochs for \civ and \siv
components are given in columns 4, 5 and 8, 9 of
Table~\ref{lab:bal_kinematics}, respectively.

The average deceleration for ``blue'' \civ BAL component is found to
be $ -1.62_{-0.05}^{+0.04}$~\cmss\ between Epochs 1 and 4, over a rest
frame time scale of 4.15 years. However, the average measured
deceleration may not be constant over time. We measure a deceleration
of magnitude $ -1.18_{-0.16}^{+0.17}$~\cmss\ between Epochs 1 and 2
which increase to $ -1.61_{-0.25}^{+0.30}$\cmss, only at $\sim$ 1.3$\sigma$ level, between Epochs 2 and 4. A similar average deceleration
is also observed for the \siv\ BAL component (see column 9 of Table
~\ref{lab:bal_kinematics}). We have also given the upper limits for
the cases where no significant acceleration is detected. It is worth
noting that based on very few cases of BAL kinematic shift reported in
the literature \citep{Gabel2003ApJ...595..120G,
  Grier2016ApJ...824..130G} the wind acceleration is not found to be
constant when multiple epochs are compared.

Next, we test the variability in strength and/or shape of BAL profile
as the highly variable optical depths may also mimic the actual
deceleration (specific to our case) signatures. For this, we perform a
$\chi^2$ test between different epoch spectra by applying a measured
velocity shift to the later epoch spectrum (see, top panel of
Fig.~\ref{fig:pixshift}). The reduced $\chi^2_{\nu}$ and corresponding
null probability $p$ of BAL profile being similar between two epochs
for \civ and \siv BAL components are listed in column $6, 7$ and $10,
11$ of Table~\ref{lab:bal_kinematics}, respectively. The higher value
of $p$ would imply that the \civ BAL profile is similar between Epoch
1 via-a-vis 2 and Epoch 2 vis-a-vis 3 but show a significant
variability between Epoch 1 and 3 and Epoch 1 and 4. Similarly, the
probability of \siv BAL profile to be different between Epoch 1 and 3
is less than 10\%, whereas, shows a significant difference between
Epoch 1 and 4 (see also, Table~\ref{lab:bal_kinematics}). A similar
trend can also be seen from the measured rest frame \civ ($W_{1549}$)
and \siv ($W_{1400}$) equivalent width between different epochs,
listed in column 6 and 7 of Table~\ref{lab:tab_sourceobs_info}. The
change in $W_{1549}$ between Epoch 1 to 2 and Epoch 2 to 3 is
significant at only $2.7\sigma$ and $2.3\sigma$ level, respectively.
For \siv BAL, the difference in $W_{1400}$ between Epoch 1 and 3 is
marginal (at $1.0\sigma$) but between Epoch 1 and 4 is significant at
$5.5\sigma$ level. Note that, both \civ and \siv absorption do not
show saturation which either imply an optically thin absorption or
partial coverage. Since, \siv optical depth is smaller than \civ and
both absorption are well detached from the corresponding emission it
indicates that \siv is optically thin even if \civ is saturated. While
the overall profile shape has not changed, the equivalent width ratio
of Si~{\sc iv}/\civ has changed by up to a factor of two (see, column
8 of Table~\ref{lab:tab_sourceobs_info}). Since, \civ may be
saturated, as a result its profile variation will be minimal compare
to the \siv when there is any change in ionizing condition (either due
to quasar luminosity change or cloud density or change in the distance
between the absorbing gas and quasar). It is also clearly evident from
Fig.~\ref{fig:xbal_vary_0911} that overall shape of the absorption
profile of the \civ and \siv BALs is quite similar over all the
epochs, except a moderate broadening of \civ BAL in Epoch 3 and 4.
CRTS $V-$mag light curve is available between MJD 53700 and 56545,
which does not show any systematic brightening/dimming of the source
though short time-scale fluctuations may be present. The mean $V-$mag
is $19.22$ with a $\sigma$ of $0.49$ mag. Thus the long time-scale
equivalent width variations may not be linked to quasar flux
variation.

The main point to understand now is the  deceleration-like
  signature shown by the ``blue'' component with very similar profile
at different epochs albeit having mild evolution in the observed rest
equivalent widths. In the following section we consider
different possibilities.

\begin{table*}
{\small 
 \centering
 \begin{minipage}{165mm}
\caption{Kinematic shift measurements for \civ and \siv BALs in J092345$+$512710.}
\label{lab:bal_kinematics}
\begin{tabular}{@{}rrc rrr rcc ccccl@{}}  
\hline
 \multicolumn{2}{c}{       } 
& \multicolumn{1}{c}{} 
& \multicolumn{4}{c}{\civ BAL}
& \multicolumn{1}{c}{}
& \multicolumn{4}{c}{\siv BAL}\\
\cmidrule{4-7}\cmidrule{9-12}
 \multicolumn{2}{c}{Spectra} 
& \multicolumn{1}{c}{$\Delta t^{\textcolor{blue}{a}}$} 
& \multicolumn{1}{c}{Vel shift}
& \multicolumn{1}{c}{Accel}
& \multicolumn{2}{c}{Shifted}
& \multicolumn{1}{c}{}
& \multicolumn{1}{c}{Vel shift}
& \multicolumn{1}{c}{Accel}
& \multicolumn{2}{c}{Shifted}
\\

\multicolumn{2}{c}{}
&\multicolumn{1}{c}{(years)}
&\multicolumn{1}{c}{(\kms)}
&\multicolumn{1}{c}{(\cmss)}
&\multicolumn{1}{c}{$\chi^2_{\nu}$}
&\multicolumn{1}{c}{$p$}
& \multicolumn{1}{c}{}
&\multicolumn{1}{c}{(\kms)}
&\multicolumn{1}{c}{(\cmss)}
&\multicolumn{1}{c}{$\chi^2_{\nu}$}
&\multicolumn{1}{c}{$p$}
\\

\multicolumn{1}{c}{(1)}
&\multicolumn{1}{c}{(2)}
&\multicolumn{1}{c}{(3)}
&\multicolumn{1}{c}{(4)}
&\multicolumn{1}{c}{(5)}
&\multicolumn{1}{c}{(6)}
&\multicolumn{1}{c}{(7)}
& \multicolumn{1}{c}{}
&\multicolumn{1}{c}{(8)}
&\multicolumn{1}{c}{(9)}
&\multicolumn{1}{c}{(10)}
&\multicolumn{1}{c}{(11)}

\\
\hline 
    SDSS-52252   & IGO-55652       & 2.95   &$  -1098   _{-151   }^{+158   } $  &$   -1.18_{-0.16}^{+0.17}$ &0.74&0.99&&  $-$                  &        $-$           &$-$&$-$  \\
    IGO-55652    & BOSS-57046 & 1.21   &$   -613   _{-95    }^{+114   } $  &$   -1.61_{-0.25}^{+0.30}$ &0.83&0.94&&  $-$                  &       $-$            &$-$&$-$ \\
    SDSS-52252   & BOSS-56607    & 3.77   &$  -2072   _{-63    }^{+62    } $  &$   -1.74_{-0.05}^{+0.05}$ &2.20&0.00&&$-1855   _{-167   }^{+272   } $&  $   -1.56_{- 0.14}^{+0.23}$ &1.16&0.10 \\ 
    SDSS-52252   & BOSS-57046   & 4.15   &$  -2126   _{-66    }^{+59    } $  &$   -1.62_{-0.05}^{+0.04}$ &1.66&0.00&&$-1491   _{-290   }^{+280   } $&  $   -1.14_{- 0.22}^{+0.21}$ &1.29&0.01 \\
    BOSS-56607   & BOSS-57046   & 0.38   &$     55   _{-33    }^{+32    } $  &$    < 0.96            $ &1.26&0.01&& $  99     _{-69    }^{+126   }$& $    < 2.07              $ &1.17&0.10  \\
\hline                                                                                                                                                      
\end{tabular}     
\newline
$^a$ Time-scale measured in the quasar rest frame.
\end{minipage}    
}

\end{table*}

\section{Discussion and Conclusions}
\label{lab:xbal_DnC}

We report on the kinematic shift of \civ and \siv BAL profiles in a
radio-loud quasar SDSS-J092345$+$512710. This quasar belongs to a rare
sub-class of X-ray bright BAL quasars with a neutral hydrogen column
density of $N_{\rm H} < 3 \times 10^{22}\ \rm cm^{-2}$ and an optical
to X-ray spectral index, $\alpha_{ox}$, of $\sim- 1.59$ which is
greater than the typical $\alpha_{ox}$ measured for soft X-ray weak
quasars, i.e., $\alpha_{ox} < -2$ \citep{Giustini2008A&A...491..425G}.
In addition, the difference between observed $\alpha_{ox}$ and
expected $\alpha_{ox}$ from the UV luminosity, i.e., $\Delta
\alpha_{ox}$, is found to be 0.04 \citep{Giustini2008A&A...491..425G}.
We detect an average  acceleration-like signature of $-1.62 \pm
^{0.04}_{0.05}$~\cmss\ and $-1.14 \pm ^{0.21}_{0.22}$~\cmss\ in \civ
and \siv BALs trough, respectively, over a rest-frame time span of
4.15 years (Table~~\ref{lab:bal_kinematics}). We do not find any
acceleration signature for the ``red'' component. Interestingly, we
find that the measured deceleration for ``blue'' \civ BAL may not be
constant over time, the rate of deceleration between Epoch 2 and 4 is
more rapid, about factor 1.4 (significant at only $1.3\sigma$ level)
higher than Epoch 1 and 2.
 
\par

In the handful of previous studies of the kinematic shift in
individual objects \citet{Vilkoviskij2001MNRAS.321....4V},
\citet{Rupke2002ApJ...570..588R} and \citet{Hall2007ApJ...665..174H}
have measured a positive kinematic shift (i.e., acceleration) of BAL
with a typical acceleration rate of $0.035 \pm 0.016$ \cmss, $0.08 \pm
0.03$ \cmss, and $\sim 0.0154 \pm 0.025$ \cmss\ respectively. The
first detection of negative kinematic shift (i.e., deceleration) was
reported by \citet{Gabel2003ApJ...595..120G} in the Seyfert galaxy NGC
3783. Using the multi-epoch observations they have found a synchronous
kinematic shift of C~{\sc iv}, Si~{\sc iv}, and  \nv\ absorption features,
while preserving the absorption profile, with a varying deceleration
rate which raises from $-0.1 \pm 0.03$~\cmss\ to $-0.25 \pm
0.05$~\cmss\ in the later interval. In addition,
\citet{Joshi2014MNRAS.442..862J} have detected relatively larger
deceleration rate of $-0.7 \pm 0.1$~\cmss\ and $-2.0 \pm
0.1$~\cmss\ of \civ BAL in two X-ray bright BAL quasars over
rest-frame time-spans of 3.11 and 2.34 years. Recently,
\citet{Grier2016ApJ...824..130G} have performed the first systematic
search for BAL acceleration using three epoch SDSS spectra of 140 BAL
quasars over timescales of 2.5$-$5.5 years and found only 3 cases, two
acceleration and one deceleration, of monolithic velocity shift
showing an overall lack of widespread BAL acceleration. They have
measured an average acceleration/deceleration rate of
$0.63^{+0.14}_{-0.13}$~\cmss, $0.54 \pm 0.04$~\cmss\, and
$-0.83^{+0.19}_{-0.24}$~\cmss, which is comparable to the present
study. Using the upper limits for \civ BAL acceleration and
deceleration in 76 BAL troughs, over a rest-frame timescales of 2.5 to
5.5 years, they show that the majority of BALs exhibit stable mean
velocities to within about 3 per cent. Interestingly, for all the
three cases they have found that the wind acceleration rate is not
constant over  time.

The observed kinematic shift in BAL can be produced due to several
reasons e.g., actual line-of-sight acceleration of a shell of material
from an intermittent outflow, directional shift in the outflow, and
changes in velocity dependent quantities such as ionization state, or
covering factor \citep{Hall2002ApJS..141..267H,
  Gabel2003ApJ...595..120G,Hall2007ApJ...665..174H}. At first, we
consider the simplest case of changing radiation energy which will
lead to decrease in the injected momentum and thus will slow down the
wind. As mentioned before using the CRTS light curve we find that our
source shows a negligible variation, less than order of half a
magnitude, over the time spanned by our observations. In addition, we
do not see any significant variation in the emission line flux which
responds to the continuum. So we conclude that change in radiative
energy/momentum may not be the primary source of deceleration.

Secondly, we consider the possibility of gravitational force for the
bulk radial deceleration of the flow. Most of the accretion disk wind
models predicts the BAL outflow at a typical distance of 0.01 pc from
the central source \citep{Murray1995ApJ...451..498M,
  Proga2000ApJ...543..686P} whereas the observations suggest that the
outflows are located at much larger distances in the range of parsecs
to several kilo-parsecs. Recently, using \siv BAL trough
\citet{Xu2018arXiv180501544X} have shown that more than 75\% of BAL
outflows are at $> 100$ pc \citep[see also,][and references
  therein]{Arav2018ApJ...857...60A}. Given the central black hole mass
of J092345$+$512710 to be $3 \times 10^{9}\ \rm M\odot$
\citep{shen2011ApJS..194...45S} and assuming the absorbing cloud is at
a typical distance of 1 pc from the central ionizing source where
gravity in mainly dominated by the black hole, at the SDSS Epoch 1 we
find that the escape velocity is much lower ($\sim 5081$~\kms) than
the average outflow speed of 15,000 \kms. It indicates that the
observed deceleration-like signature of the outflow is very unlikely
caused by the deceleration of continual flow due to gravitational
force. Simultaneous coverage of different ions at high signal to noise
and spectral resolution is needed to constrain the distance of the
absorbing gas from the quasar.

Alternatively, if we consider the absorbing
clouds to be at larger distances, it is quite plausible that the
outflowing gas may interact with the ambient material in the host
galaxy which in turn may cause the deceleration of BAL winds
\citep[see,][]{Leighly2014ApJ...788..123L}. In such scenario the
interaction will also change the ionization and thermal state of the
gas thereby introducing profile variation. This is not evident from
the observed BAL profiles. In view of the fact that majority of BALs
are stable within 3 percent of their mean velocity
\citet{Grier2016ApJ...824..130G} argue that BAL cloud may not have
traveled sufficiently far to interact with the ambient medium.
However, more such examples of BAL deceleration will be crucial to
test this scenario.

In disk wind models the BAL profiles are not only produced in a steady
smooth wind \citep{Murray1995ApJ...451..498M}, but also in a unsteady
clumpy flows \citep{Proga2000ApJ...543..686P,Proga2004ApJ...616..688P}
and magnetically confined disk wind
\citep{Arav1994ApJ...427..700A,deKool1995ApJ...455..448D}. In case of
steady wind whose density and velocity as a function of radius is
governed by force equation (balance between radiative acceleration and
gravity) and mass and momentum conservation at all radial distances
`$r$', it is known that parameters such as the initial injection
position and velocity of the wind and mass outflow rate will alter
velocity and density at a radial distance from quasar (for example,
see the basic set of equations given in Section 2.4.2 of
\citealt{Borguet2010A&A...515A..22B}). Therefore, even if the quasar
luminosity does not change, any time variation in the initial
condition of the wind (launch radius, initial velocity, and mass
outflow rate) can lead to acceleration or deceleration signatures (see
Section 4.1 of \citealt{Grier2016ApJ...824..130G}). It may be noted
that, absorption profile change introduced by the radial velocity
profile change and associated density profile change due to continuity
equation will also produce ionization change effects (even when
radiation field remains constant). Therefore, while producing velocity
shift one will have independent constraints from the equivalent width
and equivalent width ratio variations.

\citet{Proga2000ApJ...543..686P} have shown that the outflowing disk
wind is self shielded and unsteady which is radially accelerated to
relatively high velocities by the UV radiation. In addition, such a
wind can be spatially inhomogeneous having velocity and time dependent
partial coverage \citep[see also,][for 3D axisymmetric disc wind
  simulations]{Dyda2018MNRAS.475.3786D,Dyda2018MNRAS.478.5006D}. This
variable covering factor may also lead to the apparent profile shape
variations \citep{Proga2012ASPC..460..171P}. Interestingly,
\citet{Waters2016MNRAS.460L..79W} have shown that the acceleration by
the radiation pressure is very efficient when flux is time-dependent,
it can lead to a large change of about factor two in the net
acceleration for a small flux variation of $\sim$ 20\%. Given the fact
that the optical flux is not a perfect tracer of the line-driving flux
and quasars may show a higher amplitude variability in UV
\citep{Welsh2011A&A...527A..15W} it is possible that observed
  deceleration-like signature can also be produced in the time
dependent disk winds.

Alternatively, the disk wind may involve many small self-shielded
clouds with low volume filling factor, driven out by radiative force
while being confined by magnetic pressure
\citep{deKool1995ApJ...455..448D}. Not only the magnetically confined
clouds can maintain a roughly constant density and ionization across
the acceleration region \citep{dekool1997ASPC..128..233D}, but also
have superthermal velocity dispersion and therefore only a few of them
can explain the observed broad and smooth BAL profiles
\citep{Bottorff2000MNRAS.316..103B}. In view of the X-ray bright
nature of J092345+512710 the BAL troughs may well favor the small
clouds scenario, instead of the homogeneous radial outflows, producing
the observed velocity shift due to acceleration or change in covering
factor and/or optical depth. Therefore, disentangling these effects
would further require variability follow-ups. Also, high signal to
noise spectra at higher spectral resolution will be important to
further constrain the outflow models.

In our BAL variability studies of X-ray bright BALQSOs till now we
have found significant profile shift in three cases.
\citet{Joshi2014MNRAS.442..862J} reported two cases of 
  deceleration-like signatures. Unlike in the present case the
deceleration is also accompanied by profile shape variation in other
two cases. However, there are few common trends we notice in all three
cases: (i) The decelerating components typically have large ejection
velocities (i.e., $> 10,000$~\kms); (ii) There is associate absorption
at low velocities without showing any signatures of acceleration; and
(iii) the optical quasar continuum has not varied appreciably.
Additionally, it is intriguing that we have not yet seen acceleration
signatures in X-ray BALs combined to the fact of high frequency of
occurrence of deceleration-like signatures in them compared to
the X-ray weak typical BALs. Confirmation of this trend in large
sample will be interesting for understanding the physical origin of
X-ray loudness (or weakness) of BAL quasars.

\section*{Acknowledgments}

We thank the anonymous referee for constructive comments
and suggestions.

This work was supported by the National Key R\&D Program of China
(2016YFA0400702, 2016YFA0400703) and the National Science Foundation
of China (11473002, 11721303, 11533001).

Funding for the Sloan Digital Sky Survey IV has been provided by the
Alfred P. Sloan Foundation, the U.S. Department of Energy Office of
Science, and the Participating Institutions. SDSS-IV acknowledges
support and resources from the Center for High-Performance Computing
at the University of Utah. The SDSS web site is www.sdss.org.

SDSS-IV is managed by the Astrophysical Research Consortium for the
Participating Institutions of the SDSS Collaboration including the
Brazilian Participation Group, the Carnegie Institution for Science,
Carnegie Mellon University, the Chilean Participation Group, the
French Participation Group, Harvard-Smithsonian Center for
Astrophysics, Instituto de Astrof\'isica de Canarias, The Johns
Hopkins University, Kavli Institute for the Physics and Mathematics of
the Universe (IPMU) / University of Tokyo, Lawrence Berkeley National
Laboratory, Leibniz Institut f\"ur Astrophysik Potsdam (AIP),
Max-Planck-Institut f\"ur Astronomie (MPIA Heidelberg),
Max-Planck-Institut f\"ur Astrophysik (MPA Garching),
Max-Planck-Institut f\"ur Extraterrestrische Physik (MPE), National
Astronomical Observatories of China, New Mexico State University, New
York University, University of Notre Dame, Observat\'ario Nacional /
MCTI, The Ohio State University, Pennsylvania State University,
Shanghai Astronomical Observatory, United Kingdom Participation Group,
Universidad Nacional Aut\'onoma de M\'exico, University of Arizona,
University of Colorado Boulder, University of Oxford, University of
Portsmouth, University of Utah, University of Virginia, University of
Washington, University of Wisconsin, Vanderbilt University, and Yale
University. 

\bibliographystyle{apj} 
\bibliography{references}
\end{document}